\documentclass[conference]{IEEEtran}
  \usepackage[pdftex]{graphicx}
  \DeclareGraphicsExtensions{.pdf,.jpeg,.png}
\usepackage{algorithm}
\usepackage{algpseudocode}

\usepackage{etoolbox}
\usepackage{tabu}
\usepackage{setspace}
\usepackage{makecell}
\usepackage{epsfig}
\usepackage[usenames, dvipsnames]{color}
\usepackage[roman]{parnotes}

\usepackage{amsthm,amsmath,amssymb}
\usepackage{amsmath}
\usepackage{mathtools}
\usepackage{breqn}
\usepackage{amsfonts}
\usepackage{amssymb}
\usepackage{amsthm}
\usepackage{mathrsfs}
\usepackage{xparse}
\usepackage{multirow}
\usepackage{chngcntr}
\usepackage{dblfloatfix}    
\usepackage{soul} 
\usepackage{cite}
\usepackage{bm}
\usepackage{footnote}
\usepackage{subcaption}
\usepackage{algorithmicx}
\usepackage{algpseudocode}
\usepackage[font={small}]{caption}

\algdef{SE}[DOWHILE]{Do}{doWhile}{\algorithmicdo}[1]{\algorithmicwhile\ #1}%

\makeatletter
\newcommand*{\algrule}[1][\algorithmicindent]{%
  \makebox[#1][l]{%
    \hspace*{.2em}
    \vrule height .75\baselineskip depth .25\baselineskip
  }
}

\newcount\ALG@printindent@tempcnta
\def\ALG@printindent{%
    \ifnum \theALG@nested>0
    \ifx\ALG@text\ALG@x@notext
    \else
    \unskip
    \ALG@printindent@tempcnta=1
    \loop
    \algrule[\csname ALG@ind@\the\ALG@printindent@tempcnta\endcsname]%
    \advance \ALG@printindent@tempcnta 1
    \ifnum \ALG@printindent@tempcnta<\numexpr\theALG@nested+1\relax
    \repeat
    \fi
    \fi
}
\patchcmd{\ALG@doentity}{\noindent\hskip\ALG@tlm}{\ALG@printindent}{}{\errmessage{failed to patch}}
\patchcmd{\ALG@doentity}{\item[]\nointerlineskip}{}{}{} 
\makeatother

\DeclareMathOperator{\tr}{tr}

\allowdisplaybreaks

\begin{document}
%
\title{An ML-assisted OTFS vs. OFDM adaptable modem}
%
\makeatletter
\def\ps@IEEEtitlepagestyle{%
  \def\@oddhead{\mycopyrightnotice}%
  \def\@oddfoot{\hbox{}\@IEEEheaderstyle\leftmark\hfil\thepage}\relax
  \def\@evenhead{\@IEEEheaderstyle\thepage\hfil\leftmark\hbox{}}\relax
  \def\@evenfoot{}%
}
\def\mycopyrightnotice{%
  \begin{minipage}{\textwidth}
  \scriptsize
  Copyright~\copyright~2023 IEEE. Personal use of this material is permitted. Permission from IEEE must be obtained for all other uses, in any current or future media, including\\reprinting/republishing this material for advertising or promotional purposes, creating new collective works, for resale or redistribution to servers or lists, or reuse of any copyrighted component of this work in other works by sending a request to pubs-permissions@ieee.org. Accepted for publication in IEEE Future Networks World Forum 2023.
  \end{minipage}
}
\author{\IEEEauthorblockN{I. Zakir Ahmed  and Hamid Sadjadpour\\}
\IEEEauthorblockA{Department of Electrical and Computer Engineering\\
University of California, Santa Cruz\\
}}

\setlength{\columnsep}{0.21 in}

\maketitle
\begin{abstract}
The Orthogonal-Time-Frequency-Space (OTFS) signaling is known to be resilient to doubly-dispersive channels, which impacts high mobility scenarios. On the other hand, the Orthogonal-Frequency-Division-Multiplexing (OFDM) waveforms enjoy the benefits of the reuse of legacy architectures, simplicity of receiver design, and low-complexity detection. Several studies that compare the performance of OFDM and OTFS have indicated mixed outcomes due to the plethora of system parameters at play beyond high-mobility conditions. In this work, we exemplify this observation using simulations and propose a deep neural network (DNN)-based adaptation scheme to switch between using either an OTFS or OFDM signal processing chain at the transmitter and receiver for optimal mean-squared-error (MSE) performance. The DNN classifier is trained to switch between the two schemes by observing the channel condition, received SNR, and modulation format. We compare the performance of the OTFS, OFDM, and the proposed switched-waveform scheme. The simulations indicate superior performance with the proposed scheme with a well-trained DNN, thus improving the MSE performance of the communication significantly.
\end{abstract}
\IEEEpeerreviewmaketitle

\section{Introduction}
The next generation of wireless communications envisions a range of new applications like holographic communications, tactile internet, autonomous vehicles, robot swarms, UAVs, high-speed trains, and space communication to name a few. This calls for aggressive requirement specifications on the key performance indicators (KPI) for the next-generation wireless communication standards. For example, in 6G, some of the KPI specifications being considered include achieving peak data rates of 1 Tbps, with a reliability of 99.99999\%, and having a network density of 10 million devices per square km. The air-interface latency being considered is 0.1 ms with dopplers up to 1000kmph \cite{tong_zhu_2021,Twelve6G}. The combination of technologies like ultra-massive Multiple-Input Multiple-Output (uMaMIMO), reconfigurable Intelligent Surfaces (RIS), and terahertz communications is seen to be a key enabler in achieving such goals \cite{6Gsurvey}. One of the important design considerations, particularly to support error-free communications at high dopplers is the choice of waveform and modulation scheme \cite{Twelve6G, SurveyWFM1}. It is known that, by designing the waveforms where the symbols are spread in the domain of selectivity, optimal performance can be achieved for perfect channel state information (CSI) assumption at the receivers \cite{Twelve6G}. Thus the Orthogonal-Frequency-Division-Multiplexing (OFDM) waveforms are susceptible to doubly-selective channel conditions that arise due to high mobility. On the other hand, the Orthogonal-Time-Frequency-Space (OTFS) waveforms spread the symbols both in time and frequency by placing them on a delay-doppler (DD) grid, ensuring that all the symbols experience “equal gain”, and hence are more amicable for high doppler scenarios \cite{Hadani1,Twelve6G}. However, the OFDM waveform offers the advantage of having low-complexity receivers with simple channel equalization and detection capabilities.

\subsection{Previous works}
A look into some of the recent literature reveals that the comparative performance between OTFS and OFDM has mixed conclusions. It is shown that OFDM when considered in conjunction with multicarrier multisymbol linear MMSE processing, performs slightly better than OTFS in terms of achievable spectral efficiency both at low and high dopplers \cite{Carmen}. A comprehensive analysis comparing OFDM and OTFS under sparse channel conditions is presented in \cite{Gaudio}. In it, the authors do not claim to establish in a fully definitive way which is the best waveform format, since such choice depends on many other features which are outside the scope of their work that include legacy, intellectual property, ease and know-how for implementation, and many other criteria. They also make the observation that OTFS achieves a better communication rate mainly because of the presence of a “per symbol” Guard Interval rather than a per-symbol Cyclic Prefix as in OFDM. This of course comes at the cost of a more complex channel estimation scheme, working on large block-wise operations. The authors also claim that OTFS is insensitive to the magnitude of the Doppler shifts, while the performance of OFDM degrades significantly even under small-to-moderate Doppler values if the number of subcarriers increases. Therefore, OTFS is effectively a good candidate for high-mobility systems in rural environments when the delay-doppler channel is sparse.\\
\indent Although performance resilience is warranted at high dopplers and at sparse channel conditions where the OTFS comes to the rescue, most of the user equipment (UE) in an urban setting often experience low-doppler, rich scattering conditions consisting of multiple reflective paths with fixed reflectors (targets). In such a scenario, the OFDM modem with efficient computational complexity typically offers a robust performance similar to or better than OTFS waveforms. This serves as one of the main motivations to enable waveform-switching modems in the next generation of wireless standards that ensure a high-quality user experience seamlessly across varying use cases. However, as claimed by the authors in \cite{Gaudio}, it is challenging to conclusively ambit the advantage of one waveform over the other given a set of operating conditions. Keeping this in mind, and partly inspired by the idea of “Gearbox PHY for 6G” discussed in \cite{GearboxPHY}, we propose an adaptive framework to switch the transmitter and the receiver signal processing chain to use either an OFDM or OTFS scheme. A DNN-based binary classifier decides on the waveform to be used within the modem by looking at the channel conditions, received SNR, and modulation alphabet size.

\subsection{Our contributions}
The main contributions of this paper include-\\
\textit{(i)} For the first time in the literature, to the best of our knowledge, we propose a waveform-switching wireless modem architecture that can employ either an OTFS or OFDM waveform format based on the observed channel condition, received SNR, and modulation alphabet size.\\
\textit{(ii)} An ML-based switching mechanism is envisioned due to the challenges in deriving a specific threshold for the aforementioned mode switching, which is an open problem. The previous works discussed above, and our simulations epitomize these challenges.\\

\indent \textit{Notations :}
The column vectors are represented as boldface small letters and matrices as boldface uppercase letters. The expectation operator is denoted as $E[\cdot]$. The multivariate circularly-symmetric normal distribution with mean $\boldsymbol{\mu}$ and covariance $\boldsymbol{\varphi}$ is denoted as $\mathcal{CN}(\boldsymbol{\mu},{\boldsymbol{\varphi}})$. The trace of a matrix $\bold{A}$ is shown as $\tr{(\bold{A})}$ and $\bold{I}_N$ represents a $N \times N$ identity matrix. The superscript $H$ indicates the Hermitian transpose. The symbol $\otimes$ represents the kronecker product, and the vectorizing operation of a matrix $\bold{A}$ is denoted as Vec$(\bold{A})$.

\section{System model}
\begin{figure}[h!]
\centering
\includegraphics[scale=0.24]{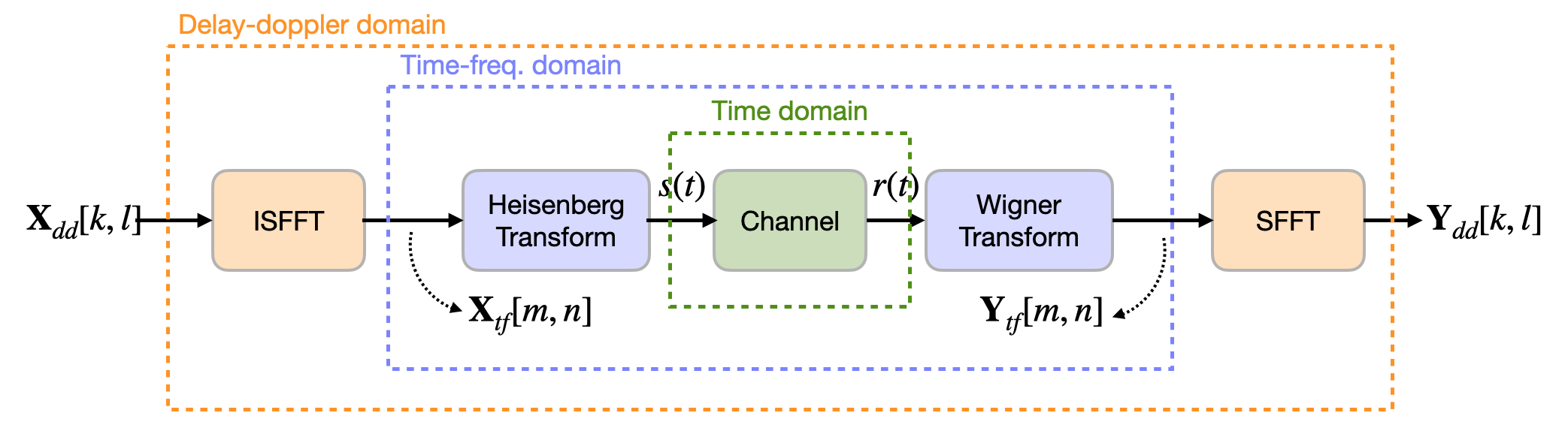}
\caption{\footnotesize  OTFS using two step approach \cite{RaviTeja1}}
\label{fig1a}
\end{figure}

In this section, we shall briefly review the OTFS system model and its relationship to OFDM \cite{RaviTeja1}. The input-output (IO) relationship considering a linear MMSE combiner at the receiver is presented. The MSE expression for the OTFS and OFDM structures is derived using the IO relationship.
The OTFS modulator places the information symbols $X_{dd}[k,l]$ along the DD grid of dimension $N \times M$ such that $k = 0,1,\cdots,N-1$ and $l = 0,1,\cdots,M-1$. The OTFS burst has a duration of $NT_0$ and occupies a bandwidth of $B = MF_0$, where $T_0$ and $F_0$ are the time and frequency resolution of the time-frequency (TF) plane of the burst under consideration such that $T_0F_0 = 1$. Hence the TF plane is characterized as $\Lambda = \Big\{ (nT_0, mF_0), n = 0,1,\cdots,N-1, m = 0,1,\cdots,M-1 \Big\}$, and the DD plane as  $\Gamma = \Big\{ (\frac{k}{NT_0}, \frac{l}{MF_0}), k = 0,1,\cdots,N-1, l = 0,1,\cdots,M-1 \Big\}$. The terms $\nu_0 = \frac{1}{NT_0}$ and $\tau_0 = \frac{1}{MF_0}$ represent the doppler shift and the delay resolution of the DD plane \cite{RaviTeja1}. An inverse symplectic finite Fourier transform (ISFFT) is performed to transform the symbols to TF representation $X_{tf}[n,m]$ which are later converted to time domain signals $s(t)$ using Heisenberg transform for transmission. 
\begin{equation}\label{eq_3a}
\begin{split}
&X_{tf}[n,m] =  \frac{1}{NM}\sum_{k=0}^{N-1} \sum_{l=0}^{M-1} X_{dd}[k,l] e^{j2 \pi (\frac{nk}{N} - \frac{ml}{M})},\\
&s(t) = \sum_{n=0}^{N-1} \sum_{m=0}^{M-1} X_{tf}[n,m] g(t - nT_0) e^{j2 \pi m F_0 (t-nT_0)},
\end{split}
\end{equation}
$g(t)$ is the unit-energy pulse of duration $T_0$. At the receiver, the received signal $r(t)$ that is corrupted by a doubly-selective channel with TF response $H(t,f)$ and AWGN noise $n(t)$ is transformed to TF representation $Y_{tf}[n,m]$ using Wigner transform, which is later converted back to DD domain using SFFT as $Y_{dd}[k,l]$ as shown in \eqref{eq_3b}. The OTFS mod-demod (modem) architecture is shown in Fig \ref{fig1a}. 
\begin{equation*}
\begin{split}
&r(t) = \int  H(t, f) S(f) e^{j 2\pi f t} dt + n(t),\\
&Y_{tf}[t,f] \triangleq \int g(t'-t)r(t') e^{-j 2\pi f(t'-t)} dt',
\end{split}
\end{equation*}
\begin{equation}\label{eq_3b}
\begin{split}
&Y_{tf}[n,m] =  Y_{tf}[t,f] \Big|_{t = nT_0, f=mF_0 },\\
&Y_{dd}[k,l] =  \frac{1}{NM}\sum_{n=0}^{N-1} \sum_{m=0}^{M-1} Y_{tf}[n,m] e^{-j2 \pi (\frac{nk}{N} - \frac{ml}{M})}.
\end{split}
\end{equation}
Here $S(f)$ is the Fourier transform of $s(t)$. The TF response $H(t,f)$ of a doubly-dispersive channel can be written as
\begin{equation}\label{eq_1}
\begin{split}
H(t,f) = \sum_{k=0}^{N_p-1}\alpha_k e^{-j 2 \pi \tau_k f} e^{-j 2 \pi \nu_k t},
\end{split}
\end{equation}
where $N_p$ is the number of multi-path components between the transmitter and the receiver, $\alpha_k, \tau_k$, and $\nu_k$ represent the gain, delay, and doppler shift associated with the $k^{th}$ path, respectively. In the discrete form, we have $H_{tf}[n,m] = H(nT_0, m F_0)$.  The complex baseband representation of the channel in the DD domain $H_{dd}[k,l] = H_{dd}(k\tau_0, l\nu_0)$ and the TF representation of the same $H_{tf}[n,m]$ are interchangeable using the SFFT/ISFFT.
\begin{figure*}[b!]
\centering
\begin{minipage}[b]{1.0\textwidth}
\begin{center}
\includegraphics[scale=0.4]{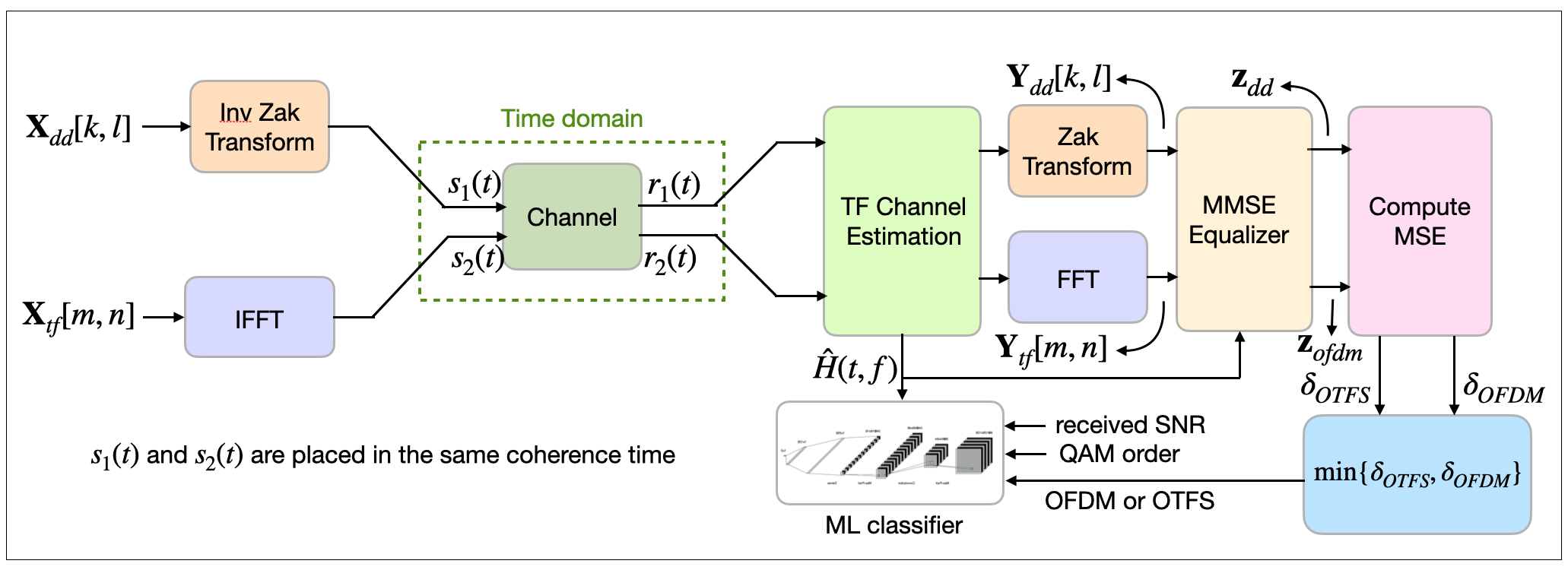}
\subcaption{Modem during ``Training".}\label{fig2a}
\end{center}
\end{minipage}
\vfill
\begin{minipage}[b]{1.0\textwidth}
\begin{center}
\includegraphics[scale=0.4]{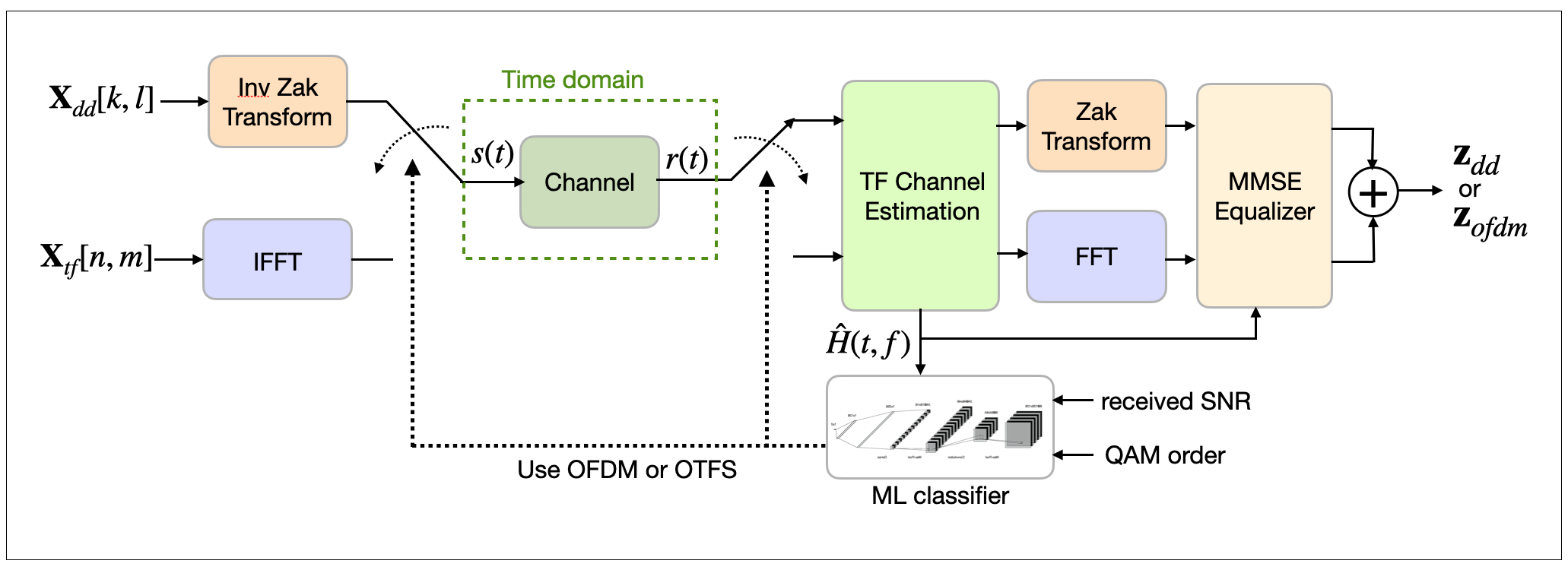}
\subcaption{Modem during normal operation.}\label{fig2b}
\end{center}
\end{minipage}
\caption{Proposed ML-assisted waveform-switching modem architecture.}\label{fig2}
\end{figure*}
\subsection{OTFS}\label{otfs_sec}
The IO relationship between the demodulated waveform $Y_{dd}[k,l]$ and the input symbols $X_{dd}[k,l]$ can be compactly written as \cite{AkbarSyed} 
\begin{equation}\label{eq_4}
\begin{split}
\bold{y}_{dd} = \sqrt p \bold{H}_{dd} \bold{x}_{dd} + \bold{n}_{dd},
\end{split}
\end{equation}
where $\bold{x}_{dd} = \text{Vec}(\bold{X}_{dd})$, $\bold{H}_{dd} = \bold{U}_{sfft}\bold{H}_{tf}\bold{U}_{\text{sfft}}^H$, $\bold{H}_{tf} = \bold{U}_{tf}^H \bold{\tilde{H}} \bold{U}_{NM}^H \bold{U}_{tf}$, and $\bold{n}_{dd} = \bold{U}_{dd}^H \bold{n}$ with $n \sim \mathcal{CN}(\bold{0},\sigma_n^2 \bold{I}_{NM})$. Here $\bold{U}_{sfft} = \bold{U}_M \otimes \bold{U}_N^H$, and $ \bold{U}_{dd} = \bold{U}_{tf} \bold{U}_{sfft}^H$. The matrix $\bold{U}_{tf}$ has dimension $NM \times NM$ whose columns are the sampled version of the TF basis function $\phi_{n,m}(t) = g(t - nT_0) e^{j2 \pi m F_0 t}$. The unitary matrix $\bold{U}_N$ is a $N \times N$ DFT matrix and $\tilde{H}[n,m] = H_{tf}[n,m]e^{\frac{j2 \pi nm}{NM}}$ \cite{AkbarSyed}. The term $p$ is the average power of the input symbols $\bold{x}_{dd}$, and $\sigma_n^2$ is the average received noise power.\\
We consider a linear MMSE equalization at the receiver such that $\bold{z}_{dd} = \bold{W}^H \bold{y}_{dd}$, where $\bold{W} = (\bold{H}_{dd} \bold{H}_{dd}^H + \lambda \bold{I}_{\text{NM}})^{-1}\bold{H}_{dd}$. Thus the MSE $\delta_{\text{otfs}}$ is evaluated as
\begin{equation}\label{eq_5}
\begin{split}
\delta_{\text{otfs}} &= \tr \Big\{ E [ (\bold{z}_{dd} - \bold{x}_{dd})(\bold{z}_{dd} - \bold{x}_{dd})^H] \Big\}.
\end{split}
\end{equation}
Alternately, the OTFS modulation can also be realized using a single-step approach using Zak transforms \cite{Saif3}. However, for evaluation of the performance of the OTFS scheme, we will consider the two-step approach discussed above. The proposed scheme can also be used with the Zak-transform-based OTFS structure without any loss of generality.
\subsection{OFDM}\label{ofdm_sec}
It can be observed from \eqref{eq_3a} that by using a rectangular pulse $g(t)$ of duration $T_0$, the inner summation over the index $m$ is reduced to an $M$-point inverse Fourier transform (IFT). Hence, a single burst of OTFS symbol $s(t)$ can be visualized as an SFFT precoding applied on $N$ consecutive independent OFDM symbols with $M$ subcarriers. To realize an OFDM scheme comparable to the OTFS the information alphabets $X_{tf}[n,m]$ are placed directly on the TF grid $\Lambda$ without having to perform the inverse SFFT on them and have IFT implemented along the columns of $X_{tf}[n,m]$ for $n=0,1,\cdots, N-1$ \cite{RaviTeja1}.
\begin{equation}\label{eq_7}
\begin{rcases}
s_n(t) &= \sum_{m=0}^{M-1} X_{n}[m] e^{j2 \pi m F_0 t},\\
Y_{n}[m] &= \int_{0}^{T} r(t) e^{-j2 \pi m F_0 t} dt.
\end{rcases}
\text{ for }n = 0, 1, \cdots, N-1
\end{equation}
Accordingly, changing the matrices in \eqref{eq_4} and using the MMSE equalization, the MSE $\delta_{\text{ofdm}}$ of the OFDM scheme can be evaluated \cite{AkbarSyed}.
\begin{figure*}[h!]
\centering
\begin{minipage}[b]{.43\textwidth}
\includegraphics[scale=0.42]{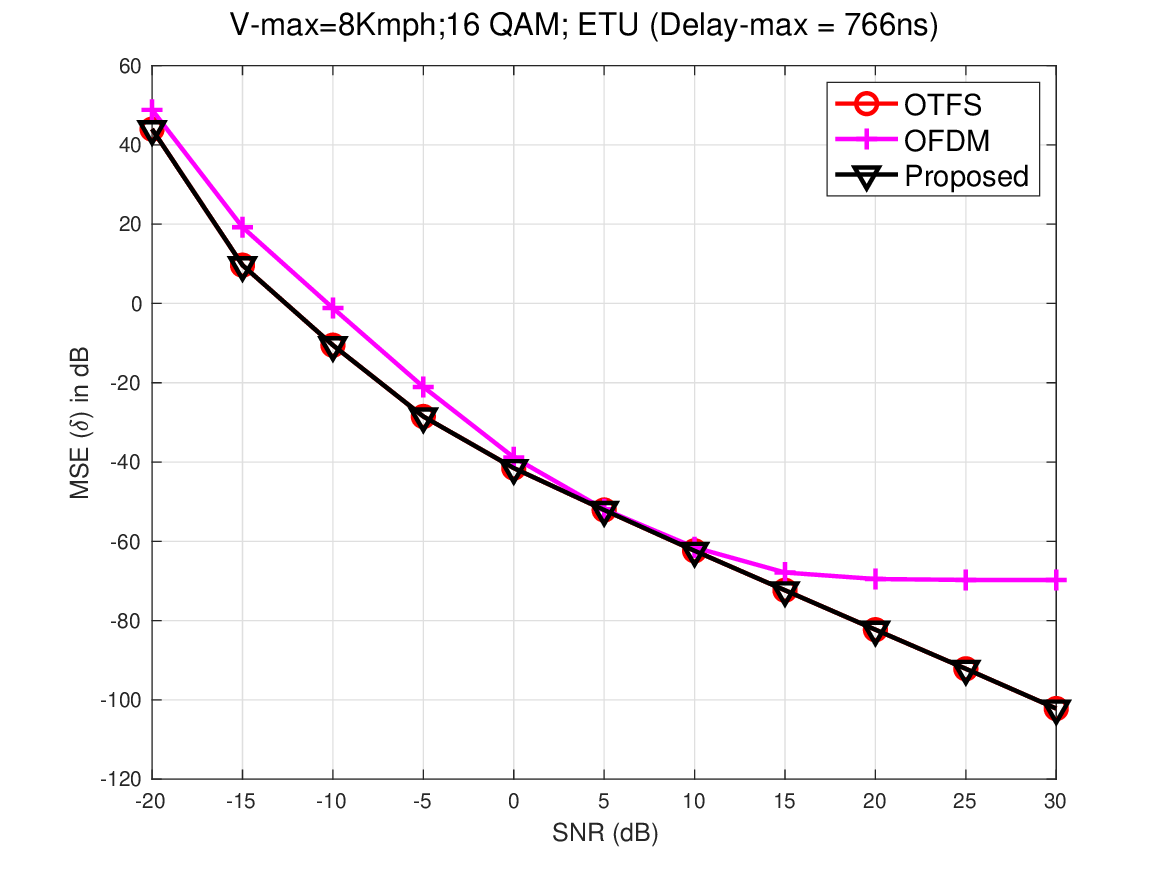}
\subcaption{Scenario-1}\label{fig5a}
\end{minipage}\qquad
\begin{minipage}[b]{.4\textwidth}
\includegraphics[scale=0.42]{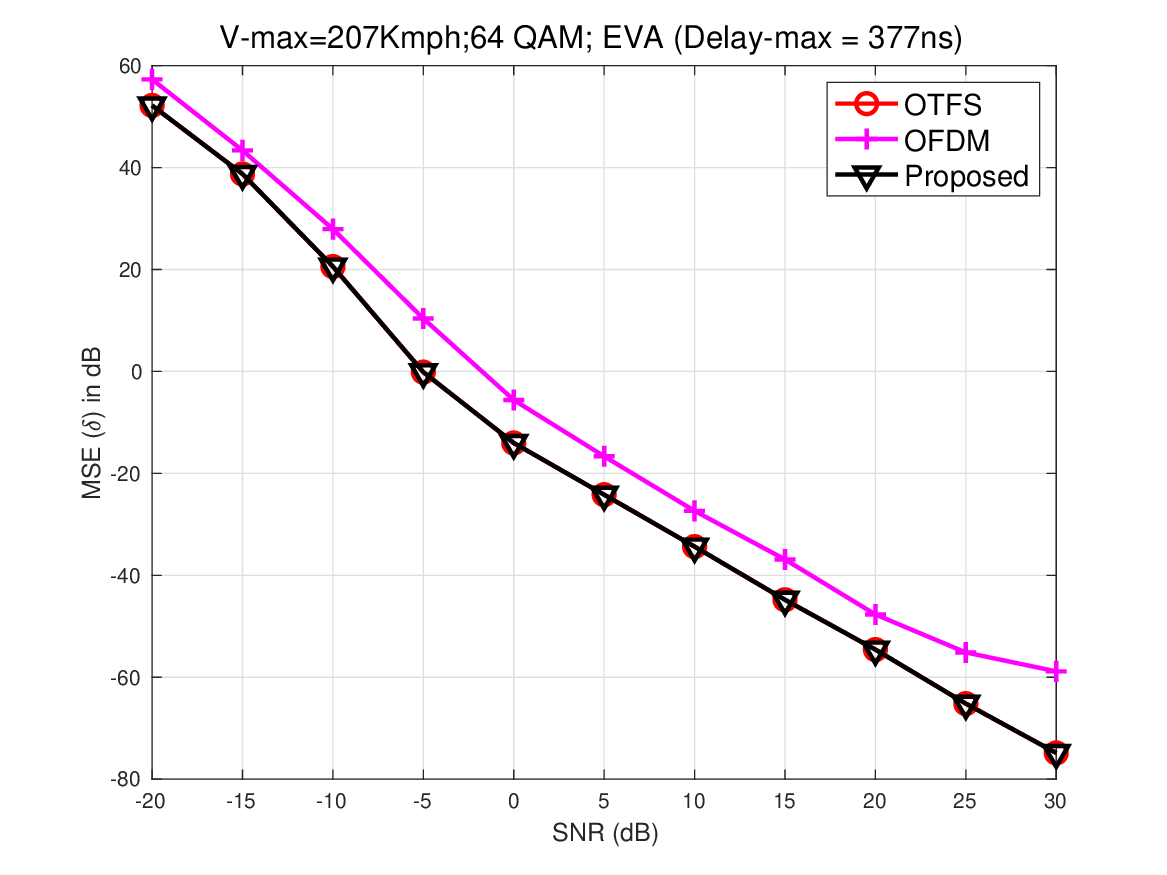}
\subcaption{Scenario-2}\label{fig5b}
\end{minipage}
\\
\bigskip
\begin{minipage}[b]{.43\textwidth}
\includegraphics[scale=0.42]{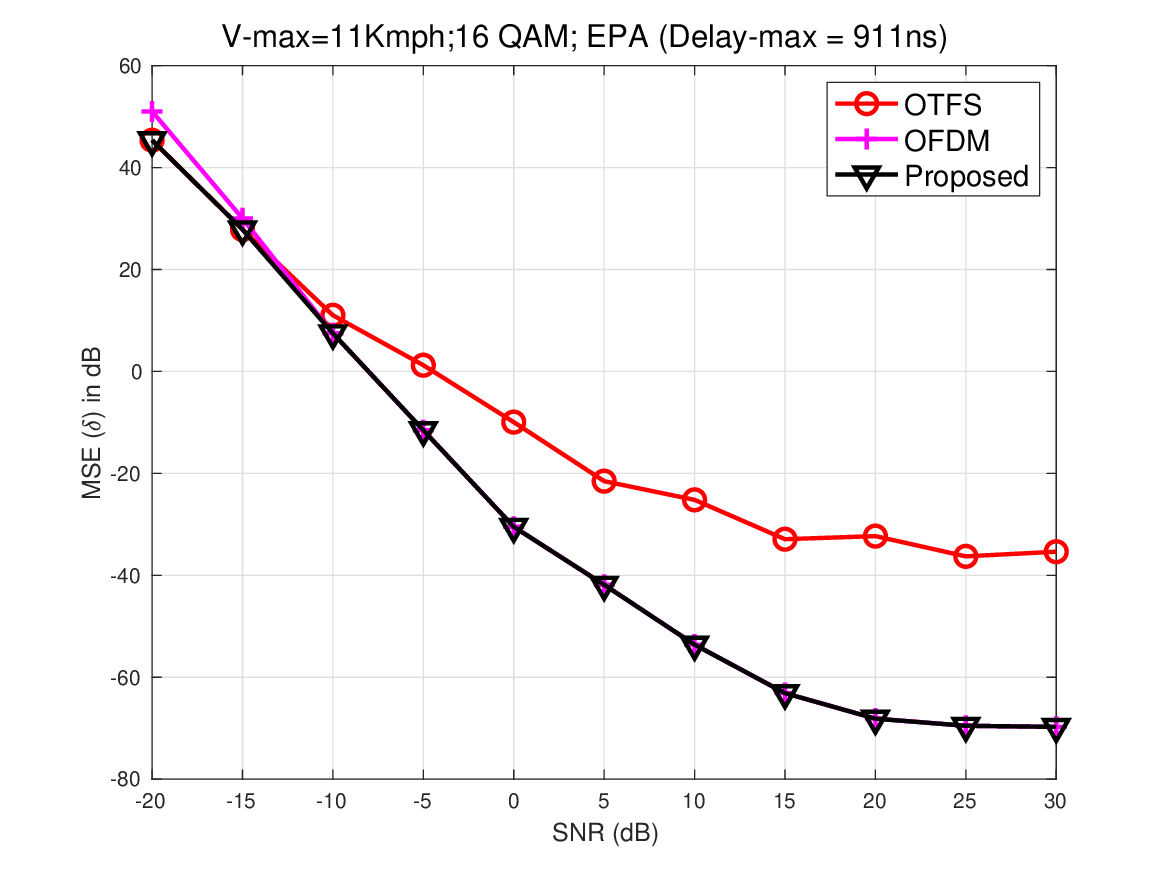}
\subcaption{Scenario-3}\label{fig5c}
\end{minipage}\qquad
\begin{minipage}[b]{.4\textwidth}
\includegraphics[scale=0.42]{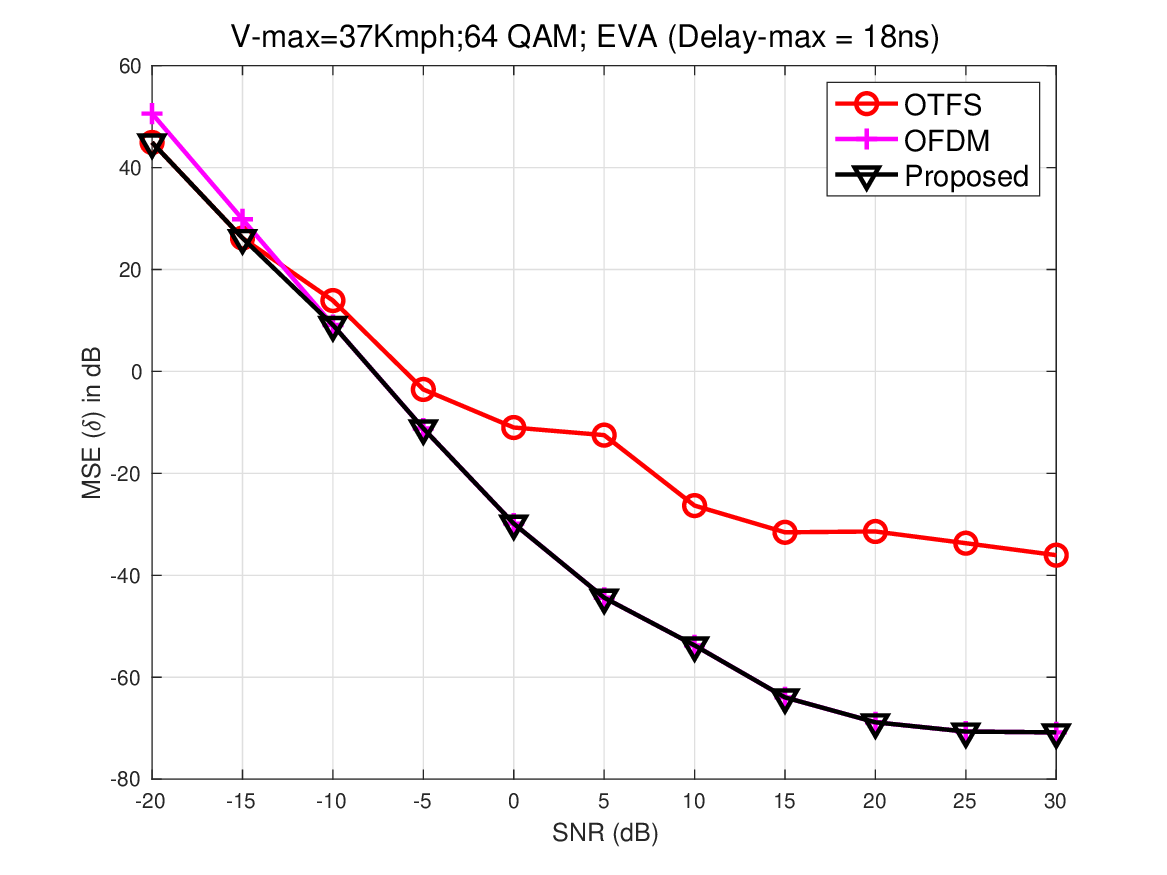}
\subcaption{Scenario-4}\label{fig5d}
\end{minipage}
\caption{\footnotesize MSE performance with OTFS, OFDM, and proposed structure.}\label{fig_128}
\end{figure*}
\begin{figure*}[b!]
\centering
\includegraphics[scale=0.38]{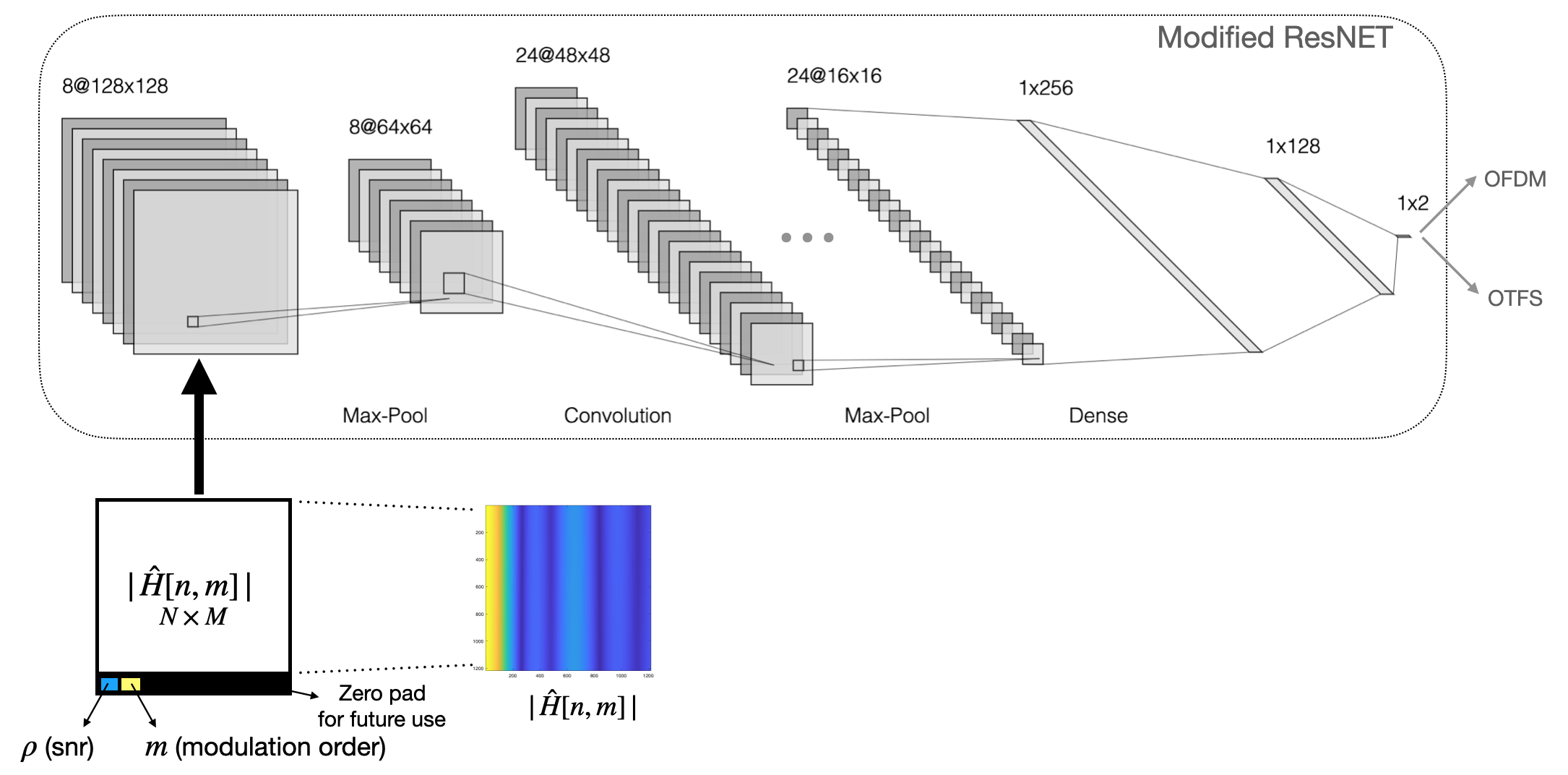}
\caption{Modified ResNET classifier}\label{fig4}
\end{figure*}
\section{Proposed modem architecture}\label{prop_arch}
In the proposed architecture, a binary ML classifier at the receiver observes the estimated TF channel $\hat{H}_{tf}[n,m]$, the estimate of the received SNR, and the QAM alphabet size to decide on the use of OTFS or OFDM waveforms in the current coherent interval. The modem operates in two modes, namely, training and realization modes, which are illustrated using Fig. \ref{fig2}. During the training phase of the modem, both OTFS and OFDM frames are transmitted within the same coherence time. There are many ways in which it can be accomplished. One possible design is to interleave the OFDM and OTFS frames. The receiver chains of the OFDM and the OTFS correspondingly extract the demodulated symbols $\bold{z}$ after channel estimation and equalization to compute the MSE $\delta_{\text{otfs}}$ and $\delta_{\text{ofdm}}$. The estimate of the actual TF channel $H_{tf}[n,m]$ is $\hat{H}_{tf}[n,m]$. A series of $p$ observations of the estimated channel $\hat{H}_{tf}[n,m]$, SNR $\rho$, QAM size $m$, and the resulting MSE $(\delta_{\text{otfs}}, \delta_{\text{ofdm}})$ with the two waveforms are used to train the ML classifier. Once the ML is trained, the proposed architecture will be able to switch between the two waveforms seamlessly based on the channel conditions, SNR, and QAM size in the normal (realization) mode. The operation of the modem during the training and realization mode is depicted in Fig \ref{fig2a} and Fig \ref{fig2b}, respectively.

\section{Simulations}\label{sims}
During the ML training, a family of $p$ channels are generated with the configurations for the Extended Vehicular-A (EVA), Extended Pedestrian-A (EPA), and Extended Typical Urban (ETU) models defined in \cite{3GPP_Channel} using \eqref{eq_1}. These models have either seven or nine paths, and the delay profiles for each path are picked from a Gaussian distribution having a mean delay equal to the RMS delay, and with variance such that the path delay doesn't exceed the maximum delay as specified in the standard \cite{3GPP_Channel}. The relative attenuation values for each of the paths are used directly from \cite{3GPP_Channel} and normalized to unity power. The Doppler shift associated with each of the paths is determined using $\nu_i = \nu_{max}cos(\theta_i)$, where $\theta_i \in \mathcal{U}[0, \pi]$ \cite{3GPP_Channel}. For each of these channels, the MSE $\delta_{\text{otfs}}$ and $\delta_{\text{ofdm}}$ are evaluated at various SNRs $(\rho) \in \mathcal{S}$ and for different M-ary QAM symbols such that  $M \in \mathcal{M}$. Here, we consider $\mathcal{S} = \{-20, -15, -10, -5, 0, 5, 10, 15, 20, 25, 30 \}$ dB, and $\mathcal{M} = \{ 4,16,64,256,1024\}$. Each of the channel $H_{tf}[n,m]$ obtained is corrupted such that ${\hat{H}_{tf}[n,m]} = H_{tf}[n,m] + \epsilon(\rho)$, where $\epsilon(\rho)$ is AWGN noise proportional to the received SNR. This simulates the channel estimation errors. This training data set is represented as $\mathcal{U} = \Big\{ \{ \hat{H}_{i}[n,m] \}_{i=1}^{p}, \{ \rho_i \}_{i=1}^{p}, \{ m_i \}_{i=1}^{p}, \{ w_i \}_{i=1}^{p} \Big\}$, Here $w_i \in \mathcal{W} = \{0, 1\}$, with $0$ indicating OTFS and $1$ indicating OFDM, whichever yields the $min(\delta_{\text{otfs}},\delta_{\text{ofdm}})$.\\
\indent The training set $\mathcal{U}$ is then used to train a modified ResNET101 binary classifier \cite{resnet_ref}. The data set $\mathcal{U}$ is formatted as an image such that each TF channel $\hat{H}_{tf}[n,m]$ is converted into an image of size $(N+1) \times M$ with each pixel color intensity representative of $|\hat{H}_{tf}[n,m]|$. The additional row is zero-padded except for the two pixels that indicate the received SNR and the M-ary QAM size. This is depicted in the Fig. \ref{fig4}. To evaluate the realization mode of the modem post training, a test set of $q$ channels are generated using the models EVA, EPA, and ETU defined in \cite{3GPP_Channel} similar to $\mathcal{U}$ such that $\mathcal{U} \cap \mathcal{T} = \emptyset$. We choose $p=1000$ and $q=100$. The results obtained for a subset of the test configurations in $\mathcal{T}$ are shown in Fig. \ref{fig5a}-\ref{fig5d}. It can be observed that the proposed architecture can switch between the two waveforms effectively in all the cases presented. The modified ResNET101 achieves a classification accuracy of $97.8$\% with our simulations.  Another general observation is that the OTFS performs better than OFDM in high Doppler scenarios with EVA models, irrespective of the modulation order, and at all SNRs. On the other hand, the OFDM performs better than OTFS at low to medium dopplers in ETV and ETU models with strong LOS and when multiple doppler shift paths are present for the same delay taps.
%
\begin{table}[http]
\begin{center}
\resizebox{0.95\columnwidth}{!}{%
\begin{tabu} to 0.5\textwidth {| l| l| }
 \hline
 \textbf{Parameters}  & \textbf{Value/Type} \\
 \hline
Frequency ($f_c$) & 4Ghz \\
\hline
Bandwidth ($B$) & 15Mhz \\
\hline
SNR ($\rho$) & $-20$ to $30$ db in steps of $5$dB \\
\hline
QAM Modulation size ($m$)  & $4,16,64,256,1024$ \\
\hline
Number of multipaths ($N_p$) & 7,9 \cite{3GPP_Channel} \\
\hline
Max. Velocity ($V_{\text{max}}$) & $10,20,150,200,250$ kmph\\
\hline 
Max delay spread ($\tau_{\text{max}}$) & $300,2300,2510$nsec  \cite{3GPP_Channel} \\
\hline
Pulse duration ($T_0$) & 9$\mu$sec\\
\hline
Subcarrier spacing ($F_0$) & $111.111$Khz \\
\hline
TF or DD grid points $(N,M)$ &  $(9, 135)$ \\
\hline
CP for OFDM &  $25$\% of $T_0$ (based on $\tau_{\text{max}}$)\\
\hline
\end{tabu}}
\caption{$\text{\footnotesize The configuration parameters used in the simulations.}$}\label{tab1}
\end{center}
\end{table}

\section{Conclusion}\label{conc}
In this paper, an ML-assisted modem architecture to switch the signal processing chains within the transmitter and the receiver to enable either OTFS or OFDM waveforms is proposed. A modified ResNET101 binary classifier is trained to identify the switching criteria that depend on the channel conditions, received SNR, and QAM alphabet size to enhance the user experience in the MSE sense. This helps in unifying the advantages of both OTFS and OFDM schemes based on the operating conditions and for a given figure of merit (FoM), which is difficult to conclusively establish, as noted in previous literature and confirmed with our simulations. Using simulation, we also established that a well-trained ResNET101 binary classifier can seamlessly switch between these two waveforms to reduce the MSE at various operating conditions. Although MSE is used as the FoM in this work to elucidate the idea, one could choose other FoMs like throughput, capacity, energy efficiency, or computational complexity. Similarly, the switching criteria can be extended beyond channel conditions, received SNR, and QAM alphabet size to encompass other system parameters.

\bibliographystyle{IEEEtran}
\bibliography{ML_assisted_BibTexFile}

\end{document}